\pdfoutput=1
\documentclass[a4paper,fleqn,usenatbib]{mnras}

\usepackage{amsmath}
\usepackage{xcolor}
\usepackage{amssymb}
\usepackage{graphicx}
\usepackage{txfonts}
\usepackage{ae,aecompl}
\usepackage[multidot]{grffile}
\usepackage[capitalise]{cleveref}
\crefname{eq:}{Eq.}{Eqs.}
\crefname{fig:}{Fig}{Figs}

\newcommand{\be}{\begin{equation}}
\newcommand{\ee}{\end{equation}}
\newcommand{\bvec}[1]{{\bf #1}}

\newcommand{\tpartial}[1]{\frac{\partial\, #1}{\partial t}}

\newcommand{\dr}[1]{\dfrac{d #1}{dr}}

\title[Gravitational instability of molecular clouds]{Gravitational instability of filamentary molecular clouds, including ambipolar diffusion}

\author[Hosseinirad et al.]{
Mohammad Hosseinirad,$^{1}$\thanks{E-mail: (m.rad@birjand.ac.ir)}
Kazem Naficy,$^{1}$
Shahram Abbassi $^{2,3}$
\and Mahmood Roshan$^{2}$
\\
$^{1}$Department of Physics, University of Birjand, PO Box 615/97175, Birjand, South Khorasan, Iran\\
$^{2}$Department of Physics, School of Sciences, Ferdowsi University of Mashhad, Mashhad, PO Box 91775-1436, Iran\\
$^{3}$School of Astronomy, Institute for Studies in Theoretical Physics and Mathematics, PO Box 19395-5531, Tehran, Iran
}
\date{Accepted XXX. Received YYY; in original form ZZZ}

\pubyear{2016}

\begin{document}
\label{firstpage}
\pagerange{\pageref{firstpage}--\pageref{lastpage}}
\maketitle
\begin{abstract}
The gravitational instability of a filamentary molecular cloud in non-ideal magnetohydrodynamics is investigated. The filament is assumed to be in hydrostatic equilibrium. We add the effect of ambipolar diffusion to the filament which is threaded by an initial uniform axial magnetic field along its axis. We write down the fluid equations in cylindrical coordinates and perform linear perturbation analysis. We integrate the resultant differential equations and then derive the numerical dispersion relation. We find that, a more efficient ambipolar diffusion leads to an enhancement of the growth of the most unstable mode, and to increase of the fragmentation scale of the filament.
\end{abstract}




\section{Introduction}
Finding a reliable description of star formation has been one of the pivotal challenges of astrophysics for decades.  Having such a model for star formation which involves many diverse physical phenomena would provide invaluable information to understand the evolution of galactic structures, star clusters, binary stars and even the formation of planets. It has been long established that stars form from collapsed dense molecular clouds cores \citep[see][]{Klessen1998,Dib2007,Padoan2014,McKee-2007}. These cores are substantially denser than their parents so they collapse faster. Nevertheless, once these fragments turn into stars they start to affect the surrounding (by radiation, winds or supernova explosions) preventing it from collapsing and forming stars. Generally it should be possible to derive a model that follows it from the scale of the largest self gravitating clouds, the giant molecular clouds (GMCs) ($\sim$ 100 pc), to the scale of protostars ($\sim 10^{-5}$ pc); while this is not possible in direct hydrodynamic simulations due to resolution limits, it can be studied approximately in analytic and semi-analytic models. 

Filamentary structures are quite commonly found in the interstellar medium. Recent high angular resolution and signal to noise of dust imaging with the Herschel Space Observatory have revealed  the filamentary nature of the dense MCs in unprecedented details \citep{Andre-2010,Molinari-2010-ID667}. Consequently it has been suggested such filamentary structure likely play a key role in the star-formation process. Filaments are also ubiquitously observed in numerical simulations of supersonic turbulence in molecular clouds and of the interstellar medium \citep[e.g.][]{Klessen1998,Dib2005,Ntormousi-2016}. These filamentary structures exhibit a large degree of sub-structures of their own (such as clumps and cores). \citet{Andre-2010,Andre-2014} proposed that turbulence-driven filaments are maybe the first stage of star formation. Given the widespread interest in the filamentary clouds structure and evolution and their key role in star formation, it is worth to examine the stability properties of the filaments in their realistic setting.

It is now accepted that magnetic field plays significant role on the dynamics of molecular clouds (MCs) and consequently on the star formation process \citep{McKee-1993,Heiles-1993,Dib2010}.  Large scale magnetic field can limit compression in interstellar shocks that form cores and filaments, while the local magnetic field within cores and filaments can prevent collapse if it is large enough \citep{Mouschovias-1976}. We know that magnetic fields are coupled with ionised particles, while the gas in the GMCs and their substructures such as filaments is mostly  neutral or partially ionised. \citet{Mestel-1956} proposed a process, ambipolar diffusion, that allows charged particles to drift relative to the neutrals, with a drag force proportional to the collision rate. Also ambipolar diffusion leads to the dissipation of magnetic field as well as to driving the system into force-free states \citep{Zweibel-1997}. Consequently, ambipolar diffusion will modify effectively the dynamical role of the magnetic field on the gas, and play a key role in the star formation process.

It has been proposed that quasi-static ambipolar diffusion is the main mechanism for pre-stellar core formation. However in this quasi-static evolution model \citep{Mouschovias-1999,Ciolek-2001} the pre-stellar life time is considerably larger than most accepted values (2--5 free fall time scales) \citep{Ward-Thompson2007}. It is now generally recognised that, due to pervasive supersonic nature of flows in GMCs, core formation is not likely to be quasi-static. In a realistic picture for star formation we need to take both the ambipolar diffusion and supersonic turbulence of the gas into account.

\citet{Gehman1} studied the linear evolution of small perturbations of self-gravitating filamentary clouds. They consider wave-like perturbations to a non-uniform filament.  Due to the non-thermal nature of interstellar gas, which is confirmed observationally, they modified their equation by adding a term that attempts to model the turbulence. They numerically determined the dispersion relation and consequently the form of perturbations in the regimes of gravitational instability. In their subsequent work, in the framework of the ideal magnetohydrodynamics (MHD) theory, they added a uniform axial magnetic field and concluded that the existence of fastest growing modes and their characteristics are highly depends on the amount of turbulence and the strength of the magnetic field \citep{Gehman2}.

The purpose of this paper is to add the effect of ambipolar diffusion in the gravitational instability of filamentary MCs with magnetic field using non-ideal MHD. The outline of this article is as follows: in \cref{sec:MHD} we present the basic MHD equations, which include the induction equation with ambipolar diffusion term. We also describe physical assumptions that are applied to the system. The unperturbed state of the system is introduced in \cref{sec:unperturbed}. In \cref{sec:linear} we perform linear perturbation analysis and then describe briefly the boundary conditions and numerical method which are used to solve the perturbed equations in \cref{sec:BCs,sec:numeric} respectively. We express the results in \cref{sec:results}. In \cref{sec:discussion} we discuss our findings and finally, we summarise our conclusions in \cref{sec:conclusion}.

\section{MHD equations with AD}\label{sec:MHD}
If the degree of ionisation in a medium is very low, the force associated with the pressure gradient of the ions and their gravitational force could be neglected in comparison with the Lorentz and ion-neutral drag force. Accordingly, the Lorentz force and the drag force exerted by the neutrals on the ions could virtually balance each other.
The equations of MHD in this so-called \emph{strong coupling approximation} \citep{Shu83} with self-gravity and AD term are
\begin{equation}
  \tpartial{\rho} + \nabla\cdot\left(\rho\bvec{u}\right) = 0,
  \label{eq:cont}
\end{equation}
\begin{equation}
  \rho\tpartial{\bvec{u}} + \rho\left(\bvec{u}\cdot\nabla\right) \bvec{u}
  + \nabla p + \rho\nabla\psi
  - \dfrac{1}{4\pi}\left(\nabla\times\bvec{B}\right)\times\bvec{B} = 0,
  \label{eq:force}
\end{equation}
\begin{equation}
\tpartial{\bvec{B}} + \nabla\times\left(\bvec{B}\times\bvec{u}\right) - 
\nabla\times\Bigg\{\bigg[\eta_A\left(\nabla\times\bvec{B}\right)\times\bvec{B}\bigg]\times\bvec{B}\Bigg\}= 0,
\label{eq:ind}
\end{equation}
\begin{equation}
  \nabla^{2}\psi = 4\pi G \rho.
  \label{eq:poisson}
\end{equation}
where $\rho$ is the neutral gas density, $\bvec{u}$ is the fluid velocity, $p$ is the pressure, $\psi$ is the gravitational potential and $\bvec{B}$ is the magnetic field strength, whereas $\eta_A$ is the AD coefficient indicated by
\begin{equation}
 \eta_A = \dfrac{1}{4\pi \gamma \rho_i \rho}\label{eq:eta}
\end{equation}
\citep{Shu-1970-ID520}, where $\rho_i$ and $\gamma$ are the ions density and the drag coefficient respectively.
\citet{Draine83} proposed the value of $\gamma = 3.5 \times 10^{13}$ cm$^3$ g$^{-1}$ s$^{-1}$.
Due to the cosmic radiation, MC are partially ionised.
\citet{Elmegreen79} showed that one could approximate the relation between ions and neutrals density as
\begin{equation}
 \rho_i = C \rho^{1/2},
\end{equation}
where $C$ is a constant equals to $C = 3 \times 10^{-16}$ cm$^{-3/2}$ g$^{1/2}$.
So \cref{eq:eta} can be recast as
\begin{equation}
 \eta_A = \dfrac{1}{4\pi \gamma C \rho^{3/2}} = \dfrac{1}{4\pi \alpha \rho^{3/2}},
 \label{eq:eta_A_with_alpha}
\end{equation}
where $\alpha$ is a substitution for $\gamma C$ and equals to $\alpha = 10.5 \times 10^{-3}$ cm$^{3/2}$ g$^{-1/2}$ s$^{-1}$. These assumptions lead to a fractional ionisation of $n_i/n_n \sim 10^{-7}$ for $n_n \sim 10^4$ cm$^{-3}$, that is completely centred within $10^{-6}$ to $10^{-8}$ of observation of MC cores. By making the equations dimensionless, $\alpha$ and $\eta_A$ become $\simeq 11.465$ and $\simeq 0.007 \rho ^{-3/2}$ respectively (see \cref{sec:AD_coef}). The aforementioned fluid equations, are closed by addition of one equation of state, which is considered here to be isothermal, i.e., 
\begin{equation}
 p=P(\rho)=c_s^2 \rho,
\end{equation}
and, where $c_s$ is the isothermal sound speed which is about 0.2 km~s$^{-1}$ for a MC with temperature of 10 K and mean molecular weight of $\mu=2.36$. The above values for the number density and sound speed, will give the value of $\sim$ 0.7 for a typical MC threaded by a magnetic field of 10 $\mu$G.
\section{Unperturbed state}\label{sec:unperturbed}
The unperturbed fluid is supposed to be in the hydrostatic equilibrium state (\bvec{u}=0). We make the assumption that the initial magnetic field doesn't have any role in stability of the filament. Therefore the density and gravitational potential profile of the filament have analytical solutions \citep{Stod63,Ostriker64} and could be written in cylindrical coordinates $(r, \Phi, z )$ as
\begin{equation}
 \rho(r)=\rho_c (1+\dfrac{r^2}{8H^2})^{-2}
 \label{eq:rho}
\end{equation}
and
\begin{equation}
 \psi(r)=2\ln(1+\dfrac{r^2}{8H^2}),
\end{equation}
where $\rho_c$ is the central density and $H$ is a radial scale height of
\begin{equation}
 H=\dfrac{1}{\sqrt{4\pi G}}\left( \dfrac{c_s}{0.2~\text{km s}^{-1}}\right) \left( \dfrac{4\times 10^{-20}\text{g cm}^{-3}}{\rho_c}\right)^{1/2} \simeq 0.035~\text{pc},
\end{equation}
whereas $G$ is the gravitational constant.
Note that although the radius of the filament extends to the infinity, the mass per unit length has a finite value of
\begin{equation}
 \dfrac{M}{l}=16.4\left(\dfrac{T}{10~\text{K}}\right)~\text{M}_{\sun}~\text{pc}^{-1}.
\end{equation}
The initial magnetic field is taken to be uniform, with an axial component \be \bvec{B_0}=B_0 \hat{z} \ee and we suppose that it does not contribute to the system equilibrium. This in turn guaranties that AD does not affect the initial steady state.
\begin{figure*}
  \centering
    \includegraphics[scale=0.75]{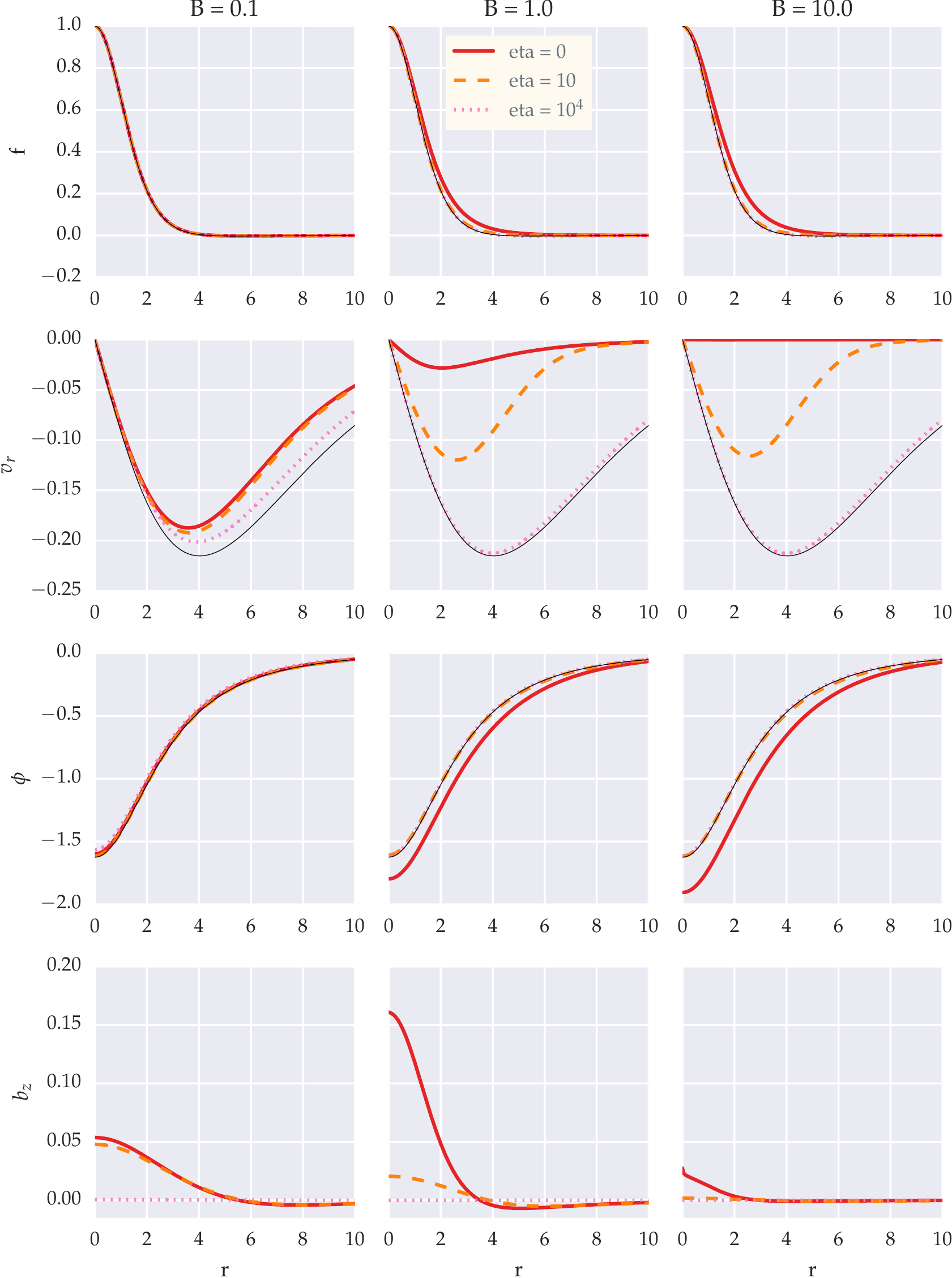}
    \caption{The eigenfunctions of the fastest growing mode in different magnetic field strength. The thick solid, dashed and dotted lines represent different $\eta_A$ values of 0, 10 and $10^4$ respectively. For comparison, the eigenfunctions when magnetic field is not present are also depicted using thin solid black lines.}
    \label{fig:eigenFunctions}
\end{figure*}

\begin{figure*}
  \centering
    \includegraphics[scale=0.7]{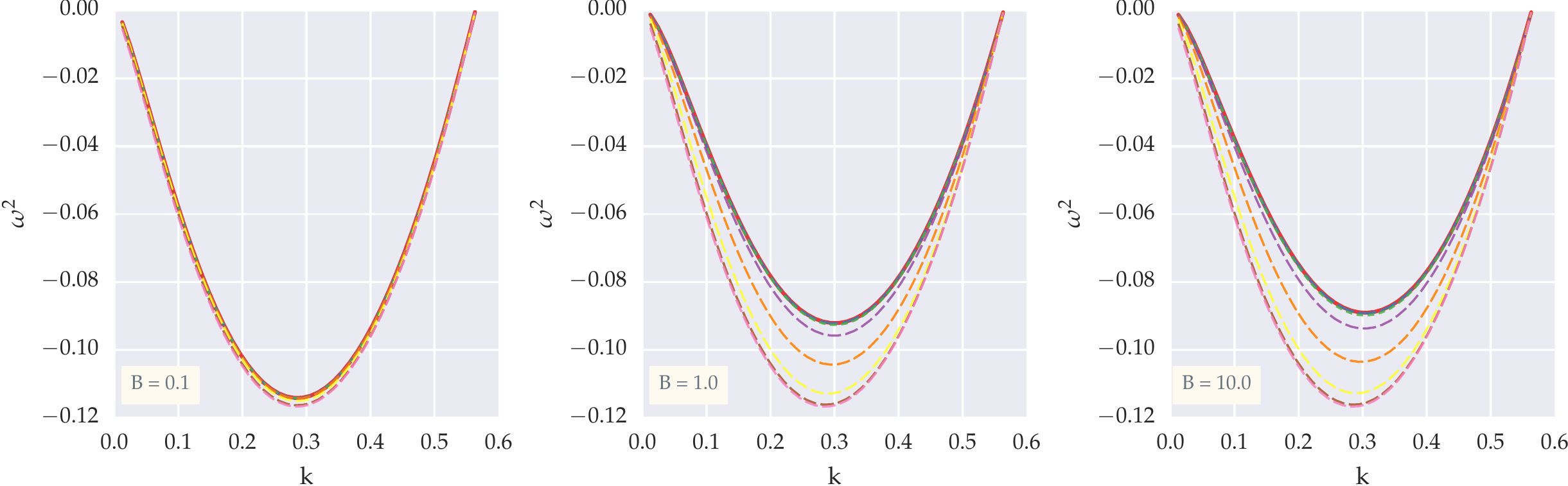}
    \caption{Dispersion relation for the filament for indicated $B$. In each figure the horizontal axis is the wave number $k$ and the vertical axis is $\omega^2$ that are normalised in the units of $(4\pi G \rho_c)^{1/2}/c_s$ and $4\pi G \rho_c$ respectively. Each dashed line demonstrates different  $\eta_A$ value from top to bottom as $0.01$, $0.1$, $1$,$10$, $10^2$, $10^3$ and $10^4$. The solid red line represents dispersion relation for the case in which $\eta_A=0$.}
    \label{fig:omega2k}
\end{figure*}
\section{Linear perturbation analysis}\label{sec:linear}
In the following, we write all the equations in dimensionless units by changing the quantities (see Appendix \ref{sec:AD_coef} for details). Unperturbed and perturbed quantities are denoted by subscript ``0'' and ``1'' respectively. We take $\eta_A$ as a constant. This will ease the calculations as well as let us to discuss the effect of AD more clearly.\footnote{The full dimensionless form of the AD coefficient is $\eta_A(r) \simeq 0.007(1+r^2/8)^{3}$.}

If we disturb the \cref{eq:cont,eq:force,eq:ind,eq:poisson} by addition of small perturbations (shown with subscript 1), they could be written to the first order as 
  \be
    \tpartial{\rho_{1}} + \nabla\rho_0\cdotp\bvec{u}_1 + \rho_0\nabla\cdotp\bvec{u}_1=0,\label{eq:cont1}
  \ee
  \be
    \rho_{0}\tpartial{\bvec{u}_{1}} + \nabla p_{1} + \rho_{0}\nabla\psi_{1}
    + \rho_{1}\nabla\psi_{0} - \left(\nabla\times\bvec{B}_1\right)\times\bvec{B}_0 = 0,\label{eq:mom}
  \ee
  \begin{equation}
\tpartial{\bvec{B}_{1}}
+ \nabla\times\left(\bvec{B}_{0}\times\bvec{u}_{1}\right) 
-\eta_A\nabla\times\Bigg\{\bigg [\left(\nabla\times\bvec{B}_1\right)\times\bvec{B}_0\bigg ]\times\bvec{B}_0\Bigg\} = 0.
\label{eq:ind1}
\end{equation}
  \be
    \nabla^{2}\psi_{1} = \rho_{1}.
  \ee 

The equation of state is assumed to be isothermal, so the pressure perturbation could be written as
\be p_1=P^{\prime}(\rho_0)\rho_1. \ee
\\
The Linear property of the equations, allows us to decompose the perturbations to their Fourier modes as
\begin{equation}
  \left(
  \begin{array}{c}
    \rho_{1}(\bvec{x}, t) \\
    \bvec{u}_{1}(\bvec{x}, t) \\
    \bvec{B}_{1}(\bvec{x}, t) \\
    \psi_{1}(\bvec{x}, t)
  \end{array}\right) =
  \Re \left[ ~ \left(
  \begin{array}{c}
    f(r) \\
    \bvec{v}(r) \\
    \bvec{b}(r) \\
    \phi(r)
  \end{array}\right) \exp{(ikz-i\omega t)} \right],
  \label{eq:rho1u1B1psi1}
\end{equation}
where $\omega$ is the angular frequency and k is the vertical wave number, whereas $f(r)$, $\bvec{v}(r)$, $\bvec{b}(r)$ and $\phi(r)$ are the amplitudes of density, velocity, magnetic field, and gravitational potential disturbances respectively. We restrict ourselves to study of the axisymmetric non-propagating evanescent unstable modes for which $i\omega$ is a real negative number. After some vector and algebraic manipulation, the complete set of equations which must be solved are yielded as
\begin{align}
-\rho_0 w + & fP''\dr{\rho_0} + P'\dr{f}+\rho_0\dr{\phi}+f\dr{\psi_0} \nonumber \\
                - & (\frac{B_0^3k^2\eta_A}{\eta_A B_0^2k^2+\Omega})\dr{b_z}-\frac{B_0^2k^2w}{\Omega(\eta_A B_0^2k^2+\Omega)} \nonumber \\ 
                + & B_0\dr{b_z}=0,\label{eq:finalODE2}
\end{align}
\be r\rho_0\dr{w}+\rho_0w+r(-\Omega^2-k^2P')f-rk^2\rho_0\phi+rw\dr{\rho_0}=0,\label{eq:finalODE3}\ee
\begin{align} 
-\eta_A & B_0^2\bigg(b_zk^2r-r(\frac{d^2b_z}{dr^2})-\dr{b_z}\bigg)-\Omega b_z r \nonumber \\
      - & B_0r\bigg(\frac{-\Omega f}{\rho_0}-k^2\frac{\phi}{\Omega}-k^2\frac{fP'}{\Omega\rho_0}+\frac{w}{\Omega\rho_0}\dr{\rho_0}\bigg)=0 \label{eq:finalODE1}\end{align}
and
\be r\frac{d^2\phi}{dr^2}+\frac{d\phi}{dr}-rk^2\phi-rf=0.\label{eq:finalODE4}\ee
where we substitute $i\omega v_r$ as $w$ and $i\omega$ as $-\Omega$. This system of four coupled ordinary differential equations (ODEs) can be converted to seven first order ODEs. In this set of equations, we treat $k$ as an eigenvalue and $\Omega$ as a constant parameter. This constitutes a two-point boundary value problem (BVP) with one eigenvalue. 
\section{Boundary conditions}\label{sec:BCs}
\Crefrange{eq:finalODE2}{eq:finalODE4} must be supplied with seven boundary conditions (BCs) to be solved. Axial symmetry demands that the radial components of velocity and all forces must vanish on the axis of symmetry. Furthermore, as the equations are linear, another boundary condition can be constrained by setting one of the dependent variables to an arbitrary value on the axis of the symmetry. We choose the density perturbation amplitude and give it the value of 1. So, we have four BCs on the filament axis that are
\begin{align}
 f&=1,\quad \frac{d\phi}{dr}=0,\quad w=0,\quad \frac{db_z}{dr}=0 \quad at\quad r=0.
 \label{boundary-0}
\end{align}
Moreover, by imposing evanescent condition, all the dependent variables and their radial derivatives have to vanish at infinity. So the remaining three BCs could be taken into account as
\begin{align}
 f&=0,\quad w=0, \quad \frac{db_z}{dr}=0 \quad at \quad r=\infty.
 \label{boundary-inf}
\end{align}
\section{Numerical method}\label{sec:numeric}
Several different methods exist to solve two-point BVPs. We use a Newton-Raphson-Kantorovich (NRK) relaxation algorithm \citep{Garaud-thesis}. The algorithm, uses the second-order finite-difference method to discretize the system of ODEs as well as BCs over a mesh grid. The resultant system of linear algebraic equations are then solved by inverting the matrix of coefficients. This procedure is iterated while an error threshold test is performed after each iteration until the solutions are converged. The reader can find more details of the algorithm in \citep{Garaud-thesis}.

We implemented this algorithm by choosing a computational interval between $r=0$ and 50. This interval was spanned with 2000 equally-spaced mesh grid.
\begin{figure}
   \includegraphics[scale=0.68]{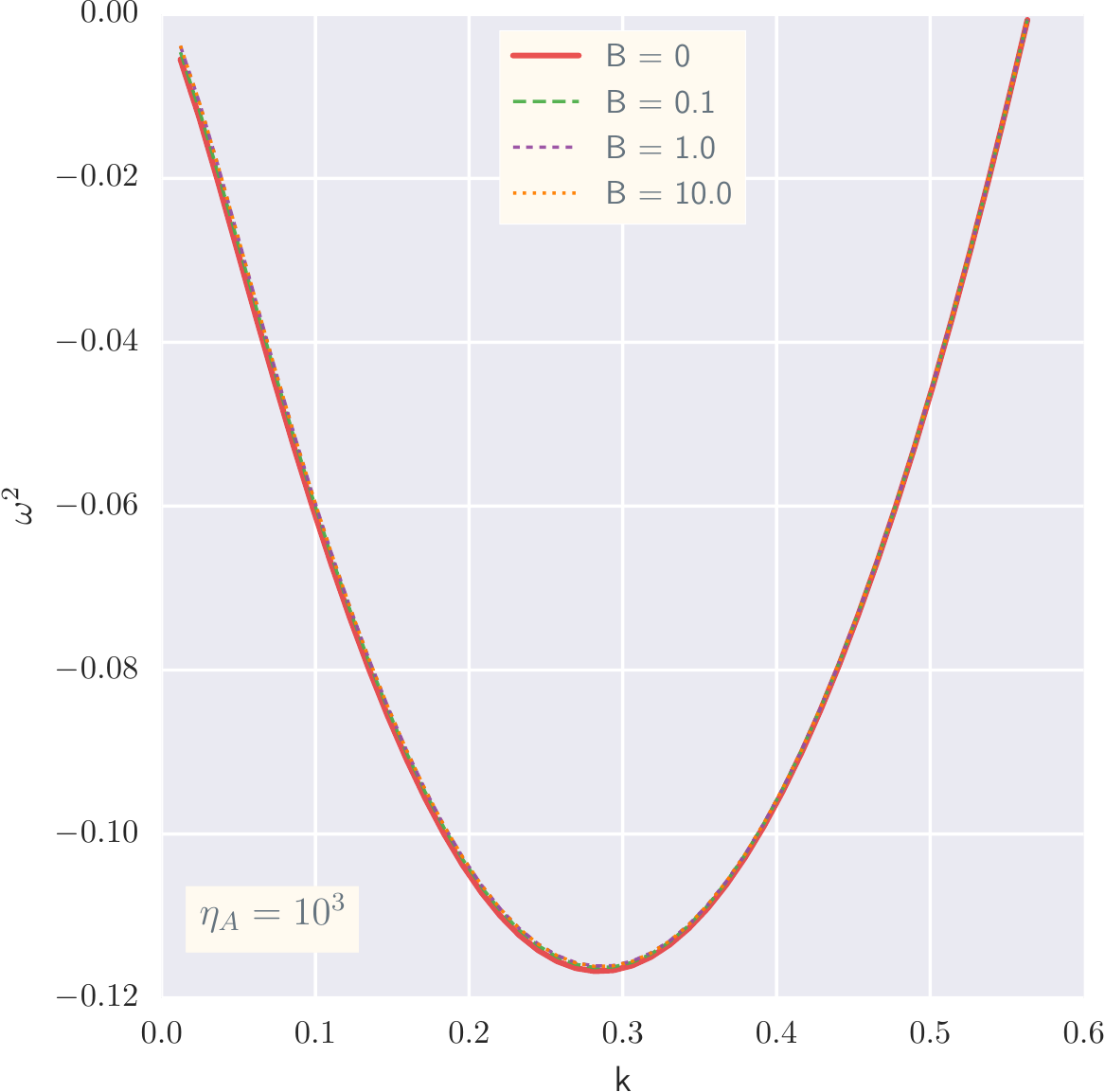}
    \caption{\label{fig:omega2kCompareEta}Comparison between $B=0.1, 1$ and $10$ in strong AD regime $\eta_A = 100$. $\omega^{2}$ and $k$ are normalised as \cref{fig:omega2k}. The solid red curve represents dispersion relation for the case in which $B=0$}  
\end{figure}
\begin{figure}
   \includegraphics[scale=0.7]{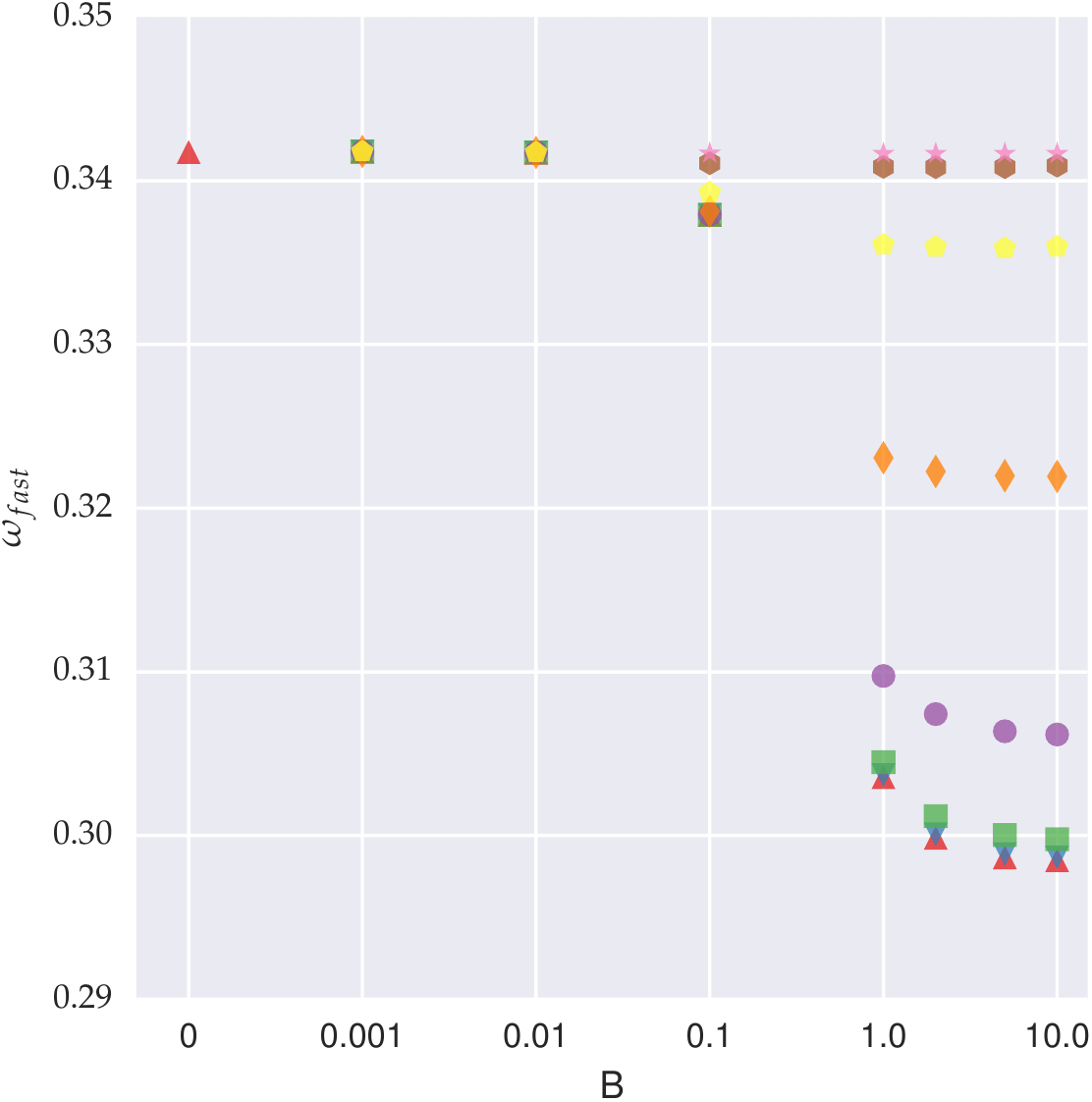}
    \caption{The fastest growing mode versus magnetic field strength. $\omega_{fast}$ and B are normalised in units of $(4\pi G \rho_c)^{1/2}$ and $(4\pi  \rho_c)^{1/2}c_s$ respectively. Markers from top to bottom demonstrate different $\eta_A$ values as $10^4$ (star), $10^3$ (hexagon), $10^2$ (pentagon),10 (diamond), 1 (circle), $10^{-1}$ (square), $10^{-2}$ (upside down triangle) and 0 (triangle). When $B=0$, $\eta_A$ has the only value of 0.}  
    \label{fig:omeFast_B}  
\end{figure}

\begin{figure}
   \includegraphics[scale=0.7]{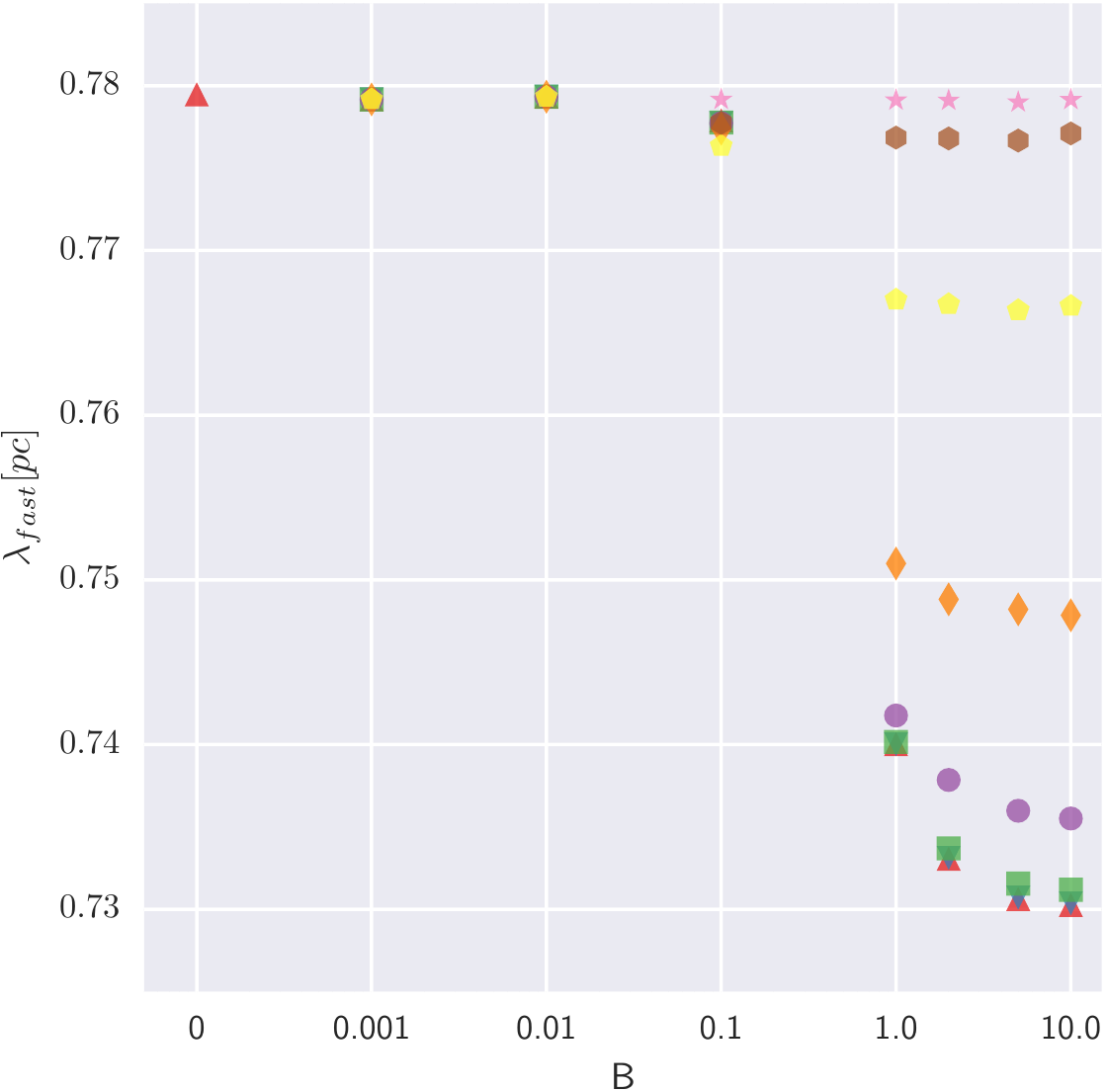}
    \caption{The length scale of fragmentation versus magnetic field strength. B is normalised as \cref{fig:omeFast_B}. B=1 corresponds with $\sim$ 14.3 $\mu$G . $\lambda_{fast}$ is in the unit of pc. Markers from top to bottom demonstrate different $\eta_A$ values as $10^4$ (star), $10^3$ (hexagon), $10^2$ (pentagon),10 (diamond), 1 (circle), $10^{-1}$ (square), $10^{-2}$ (upside down triangle) and 0 (triangle). When $B=0$, $\eta_A$ has the only value of 0.}  
    \label{fig:lambdaMin_B}  
\end{figure}

\begin{figure*}
   \includegraphics[scale=0.75]{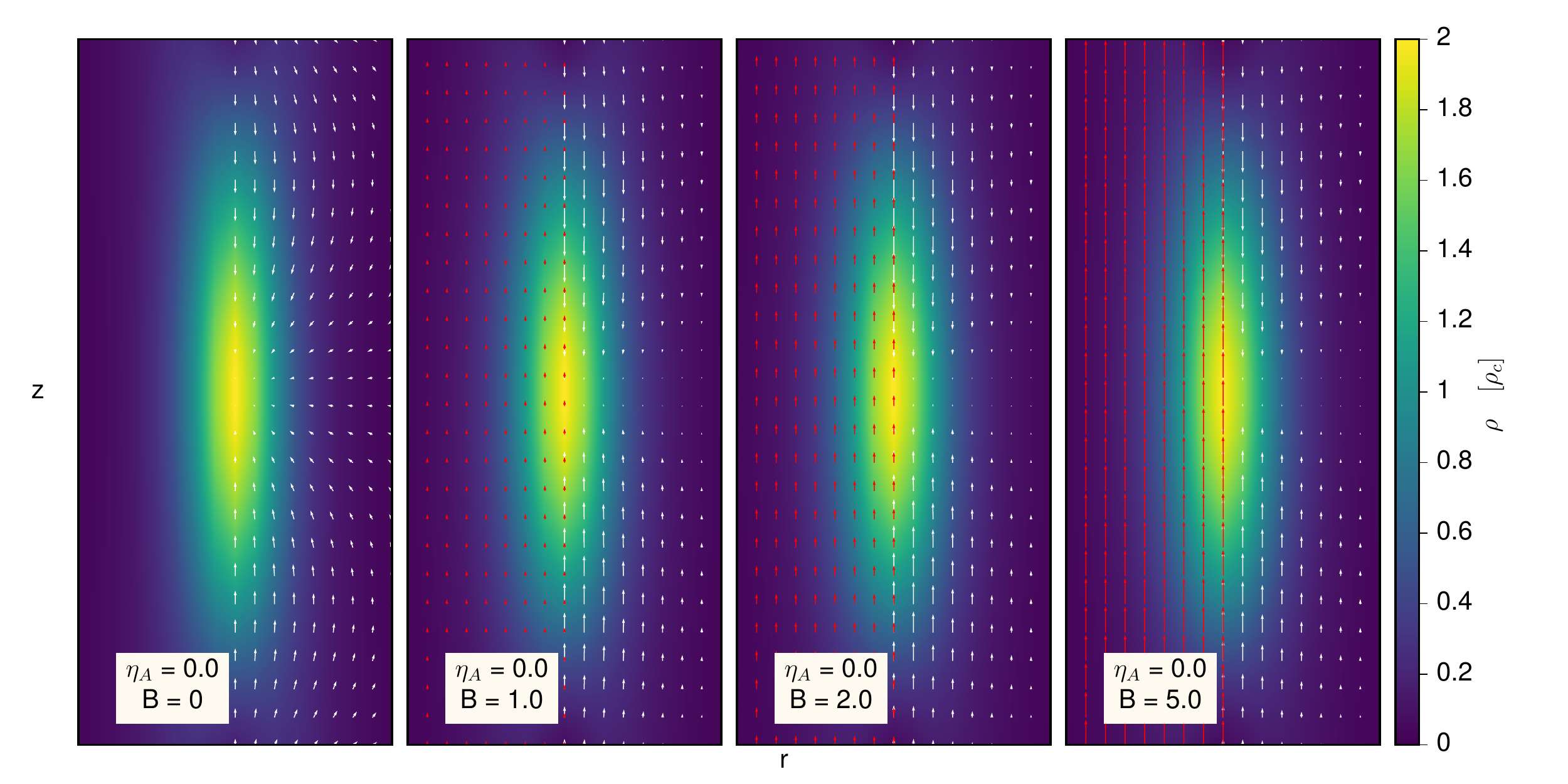}
    \caption{Cross-section of the filament \emph{without} AD effect in different magnetic field strength (from left to right B = 0, 1, 2 and 5). White and red (when present) arrows indicate total velocity and magnetic vector field overlaid on background of total density for the fastest growing mode.}  
    \label{fig:u_field}  
\end{figure*}

\begin{figure*}
   \includegraphics[scale=0.75]{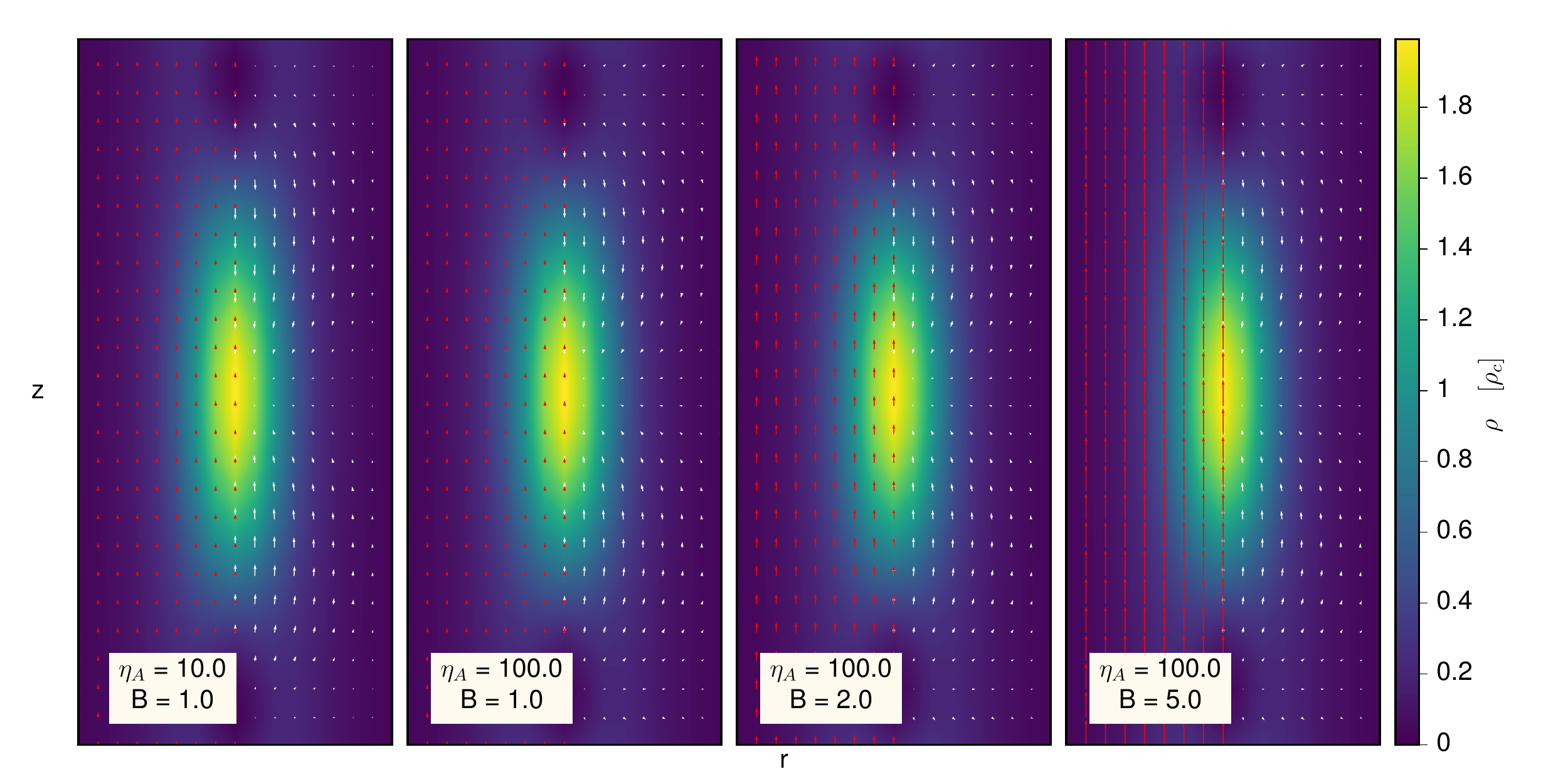}
    \caption{Cross-section of the filament \emph{with} AD effect ($\eta_A$=10 or 100) in different magnetic field strength (from left to right B = 1, 1, 2 and 5). White and red (when present) arrows indicate total velocity and magnetic vector field overlaid on background of total density for the fastest growing mode.}  
    \label{fig:u_field_big_eta}  
\end{figure*}
\section{Results}\label{sec:results}
By solving the system of ODEs, a numerical dispersion relation for the system could be computed. At first, the results of \citet{Gehman2} for different magnetic field strength of $B$, are successfully re-produced by using an approach like theirs. These results are used as a guess for the new system of ODEs which includes the effect of AD . We examined, various values for the magnetic field strength $B$ and AD coefficient $\eta_A$ to comprehend how the dispersion relation (specially the fastest growing mode) could be affected.

There exist four eigenfunctions $f$, $\phi$, $v_r$, $b_z$ related to each point on the dispersion relation curve when the AD term is added (without AD, $b_z$ as well as its derivative can be obtained algebraically, so the number of eigenfunctions is limited to three). In \cref{fig:eigenFunctions}, the eigenfunctions of the fastest growing mode are depicted for B=0.1, B=1 and B=10 each in three diffusivity regimes i.e. $\eta_A=0$, 10 and $10^4$. The computed eigenfunctions when the magnetic field and in consequence AD are absent (pure Jeans instability) are also represented for comparison. From this figure one can see that the amplitude of density perturbation $f$ has generally the same shape in all of B and $\eta_A$ values. Just in strong magnetic field, it's a little wider which by addition of AD it tends to come back to the state in which no magnetic field is present. The $v_r$ eigenfunction is more prone to change with $B$ alteration. Strong magnetic field is able to effectively damp the radial velocity perturbation. This is expected since stronger magnetic field in the $z$ direction will lead to stronger magnetic force against the radial motion of the ions. 

Here, AD  is also able to return the radial velocity to the case of no magnetic field more effectively. The treatment of $\phi$ gravitational potential eigenfunction is very much like that of density one. The AD influence is again to revert the system modes to the original state. In other words, AD suppresses the magnetic field effects and forces the system to respond against axisymmetric perturbations like a system in which there is no magnetic field. This fact is somehow surprising since one may naturally expect that increasing the AD coefficient $\eta_A$ would highlight the existence of the magnetic field in the system instead of hiding its presence. Mathematically one may conclude that large $\eta_A$ makes the $\left(\nabla\times\bvec{B}_1\right)\times\bvec{B}_0$ term in \cref{eq:ind1} small compared to other terms. This point gets more clear by noting the respond of $b_z$ to changes in $\eta_A$ in the bottom panels of \cref{fig:eigenFunctions}. It is evident that increasing the $\eta_A$ coefficient strongly damps $b_z$ and forces it to vanish. In this case, the last term in \cref{eq:ind1} would be straightforwardly very small. We will show that increasing $\eta_A$ apparently hides the magnetic field effects on the eigenfunctions, but increases the growth rate of unstable perturbations. Furthermore, the perturbation of the gravitational potential, i.e. $\phi(r)$, has also been illustrated in \cref{fig:eigenFunctions}. As expected, $\phi(r)$ similar to $f(r)$ is not too sensitive to the AD coefficient.

\cref{fig:omega2k} shows the dispersion relations for a sample of three selected magnetic field strengths, namely $B=0.1$, $B=1$ and $B=10$. When $\eta_A=0$, the magnetic field is the only agent that affects on propagation of disturbances by improving the stability of the filament. In these cases our results agree very well with previous studies \citep[see e.g.][]{Chandra53,Ostriker64,Nagasawa1987,Gehman2}. 

However, the addition of AD to the physics of the problem, changes the response of the filament to the disturbances. As expected, in the weakest magnetic field (left panel in \cref{fig:omega2k}), AD is not sufficiently efficient to change the shape of the dispersion relation even in very strong diffusion regime, just the growth rate of the fastest growing mode (i.e the largest $|\omega^2|$) has increased fairly and also shifted towards smaller wave number. For a stronger magnetic field $B=1$ (central panel), lower $\eta_A$ values, i.e. $0.01$ and $0.1$, are still almost ineffective. But for a larger value of $\eta_A\geq 1$, the growth rate of the unstable modes increases in a tangible manner. More specifically the growth rate of the most unstable mode increases significantly. This effect is more pronounced when $\eta_A=10$ and totally recognisable for $\eta_A=10^2$, $10^3$ and $10^4$. It is also worth mentioning that AD increases the most unstable wavelength but keeps the spectrum of the  unstable wavelengths constant. In other words AD does not extend the instability interval. This interval in \cref{fig:omega2k} is $0<k<0.57$. In this sense, one may conclude that AD does not make the system more unstable, and just increases the instability growth rate. 

The right panel in \cref{fig:omega2k} illustrates this situation when $B=10$. In this case, the previous story is repeated. Though when diffusion is absent the fastest growing mode is a few smaller, as $\eta_A$ gets larger, the fastest growing mode becomes smaller. Another point that one could see simply is that the critical wave number remains constant regardless of the $B$ and $\eta_A$ values.

It is instructive to mention that \citet{Gehman2} showed that a magnetic field stronger than $B\simeq 2$ is not able to further destabilise the filament. \cref{fig:omega2kCompareEta} depicts that in an inverse picture, this is also the case for very large $\eta_A$ quantities. (see also \cref{fig:omeFast_B}).

To elaborate on the dependency of the fastest growing mode on $B$ and $\eta_A$, a high-order polynomial is fitted on the calculated dispersion relation points. \cref{fig:omeFast_B} represents the result. For magnetic fields $B\leq 0.01$, even very large $\eta_A$ values are not able to change the picture. The fastest growing mode remains unchanged. When $B=0.1$, scaling up AD makes a little reduction in the fastest growing mode. In stronger magnetic field regime ($B\geq 1$), increasing AD makes a sharp decline in the fastest growing mode values. Here, it's clearly visible that the stabilising effect of magnetic field could be negated by AD.

Gravitationally unstable MC are fragmented most probably at the most rapid propagating mode. This means it's the wavelength of the fastest growing mode ($\lambda_{fast}$) which signifies a value for fragmentation length scale of the filament. Axial magnetic field could reduce this length scale by moving $k_{fast}=2\pi/\lambda_{fast}$ to some larger values \citep{Gehman2}. \cref{fig:lambdaMin_B} shows that this phenomenon could be weakened by the addition of AD effect. As for an isothermal filament fragmentation mass scale is M$_{frag}=8\pi \lambda_{fast}$ \citep{Gehman2}, AD could also increases this mass scale. 

\cref{fig:u_field} demonstrates cross-section of the filament for the fastest growing mode when AD is not present. The overall pattern of velocity and total (initial + perturbed) magnetic field (when existing) are overlaid on the total density background. The axial structure of magnetic field is preserved even in the weak magnetic field. For velocity, it's a little different. It tends to twist toward the centre of filament as the distance increases. Rising B value, attenuates this treatment and makes velocity field to lay along the filament vertical axis. \cref{fig:u_field_big_eta} is similar to its previous counterpart. It shows the situation when AD is contributing to the structure of the filament. AD can restore the twisted pattern of velocity field observed in absence of magnetic field. The axial shape of magnetic field is still remained unchanged.
\section{Discussion}\label{sec:discussion}
\subsection{The time and length scale of fragmentation}
Pertaining to the fastest growing mode, two important scales are the time-scale and length scale of fragmentation. By increasing the AD coefficient, the coupling between neutrals and ions becomes more and more fragile. This means that the support by magnetic field becomes less and less efficient until it becomes entirely inefficient and neutrals are no longer tied to the magnetic field lines at all. In other words, existence of the AD, moves the system towards Jeans instability. This fact can be seen in \cref{fig:omeFast_B,fig:lambdaMin_B}. \cref{fig:omeFast_B} shows that for all the magnetic field strengths, increasing more and more the AD effect leads all the fastest growth rates to converge to a maximum value around
\begin{equation}
|\omega|^2_{\text{max}}=0.117
\end{equation}
which is equivalent to a minimum growth time-scale of
\begin{equation}\label{eq:t_growth_min}
 \tau_{\text{min}}\simeq 3.1~{\text{Myr}}.
\end{equation}
One should notice that by \cref{eq:t_growth_min}, the minimum growth time-scale is roughly comparable or even greater than the typical life-time of MCs harbouring these filament \citep[e.g.][]{Ward-Thompson2007}. As it's pointed out in the Introduction, the turbulent flow in the interstellar medium and GMCs, could accelerate the diffusion of the magnetic field \citep{Fatuzzo2002,Zweibel2002,Kim2002,Heitsch2004,Li2012}. \citet{Kudoh2011} in a three-dimensional simulation of sub-critical sheet-like MC showed that the supersonic velocity perturbation could accelerate the AD time by an order of magnitude. In a very recent work, \citet{2016arXiv160906879B} have shown that by addition of a decaying turbulent velocity field, the AD collapse time of a filament could be decreased. It's worth noting that \citet{Gehman1,Gehman2} also derived the dispersion relations of filaments with equations of state that are softer than isothermal as a representative for turbulent fluid and showed that it can decrease the minimum growth time.

Moreover, by \cref{fig:lambdaMin_B} one could simply find that there is also a maximum value for the most unstable wavelength around
\begin{equation}
  (\lambda_{fast})_{\text{max}}=0.779~{\text{pc}}.
\end{equation}
This length scale indeed is the most unstable wavelength for the Jeans instability and certainly does not depend on the magnetic field strength as it was explained earlier for the growth rate.

It should be noticed that \citet{Ciolek2006} carried out a linear analysis to study the formation of protostellar cores in a magnetised infinite thin layer in the presence of the AD. The layer is confined within external pressure. They pointed out that for a given value of the AD coefficient, there is a maximum for the fragmentation length scale around the critical mass-to-flux ratio for collapse and increasing or decreasing of the magnetic field both yield a smaller fragmentation length scale, but a more strength magnetic field reduces the fastest growing mode; see their Fig. 2. In our work, increasing the magnetic field yields a smaller fragmentation length scale and vice versa. It should be stressed again that this is also the case in \citet{Ciolek2006} for a specific interval of mass-to-flux ratio. Furthermore as in \citet{Ciolek2006}, increasing of the magnetic field, reduces the fastest growing mode. However one should note that \citet{Ciolek2006} used a totally different geometry and naturally the response of the systems to global perturbations is , in principle, different. In other words, the discrepancy in alteration of the fragmentation length scale would be due to the different geometries.
\subsection{The effect of initial magnetic field configuration on the stability of the filament}
In this paper we considered a uniform magnetic field along the major axis of the filament as the background static magnetic field and demonstrated that the magnetic field can improve the stability of the filamentary MC until its effect is saturated \citep[see also][]{Nagasawa1987,Gehman2}. \citet{Stod63,1993PASJ...45..551N} made different assumption such that the equilibrium filament is supported by the same ratio of the magnetic field pressure to the gas pressure. This leads to an initial magnetic field which has a radial dependence proportional to $\rho_0(r)^{1/2}$. With this different configuration, contrary to our finding, the magnetic field makes the system more unstable both by increasing the growth rate of the most unstable mode and by decreasing the critical wavelength. In other work \citet{Nagasawa1987} investigated the effect of external pressure exerted on the filament boundary by a hot tenuous medium. For this kind of filament, like our case the magnetic field decreases the growth rate of the most unstable mode, but in contrast of our case in which the filament is not underwent external pressure and the critical wavelength remains unchanged, the magnetic field can increase the critical wavelength.
\subsection{The clumps spacing in filamentary MCs}
Filamentary MCs harbour clumps with typical size of 0.3--3 pc \citep[see for a review][]{2007ARA&A..45..339B} which are often found to be regularly spaced along the filaments. So it comes to mind that this equally spacing could be happened somehow by a type of instability with a particular length scale. Filaments are prone to the gravitational and MHD driven instabilities. Deformation of a cylindrical fluid surface due to the magnetic field can trigger a type of MHD instability sometimes called ``sausage`` instability in which a periodic pattern of dense regions is formed. Applying the sausage as well as the gravitational instability to the filamentary MCs, many authors attempted to predict the characteristic distance between the clumps \citep{2010A&A...520A.102M,2010ApJ...719L.185J,2011ApJ...735...64W,2012A&A...542A.101M,2014MNRAS.439.3275W,2016MNRAS.456.2041C,2016MNRAS.463..146H,2016ApJS..226....9W,2016A&A...592A..21F} by using the wavelength of the most unstable mode. The estimated characteristic space between identified clumps reported by these works, are almost in accordance with the theoretical prediction. In theory, in each unstable mode, the space between clump centres is the wavelength of that unstable mode. As the fastest growing mode predominantly determines the ongoing fragmentation properties of the filament, the space between clump centres should be determined by it's wavelength. Our results show that including the effect of AD can increase this characteristic length scale at most $\sim$ 6 per cent compared with when no AD is included. In a recent work, it is shown by \citet{2016arXiv160400378G} that geometrical bending of a filament can also result in formation of regularly spaced clumps along it. In this case, the characteristic distance of the clumps is directly determined by initial form of perturbation. Therefore, more theoretical investigations are needed to shed light on the problem.
\section{Conclusion}\label{sec:conclusion}
Perturbation analysis is a powerful tool for exploring different kinds of instability in fluid dynamics. In this work, we investigate the effect of AD on gravitational stability of a magnetised filamentary MC with an isothermal equation of state. We compute numerically the dispersion relation by linearising the governing equations and performing global perturbation analysis.  By including AD, we demonstrate that in the low magnetic strength regime, applying non-ideal MHD to the system could not alter its stability, whereas in the highly magnetised one, introduction of AD could destabilise it. It is worthy to mention that AD also destabilises the weakly ionised accretion disks, for example see \citet{Kunz2004} where the effects of AD on the magnetorotational instability have been studied.

Moreover, we find that in the highly magnetised regime, the length scale of the fragmentation $\lambda_{fast}$, which is inversely related to the $k_{fast}$, could be increased at most $\sim$ 6 per cent. This in turn means the fragmentation mass scale must be increased.

As a final remark we mention that, in this paper, we consider the global stability of an isothermal filament. On the other hand, recent observations have shown that interstellar filaments are not isothermal configurations \citep{Palmeirim2013}. Therefore a more constructive study can be done by assuming a general density profile, and not necessarily an isothermal one, and investigating the fragmentation in the presence of AD. For such a study in the absence of magnetic field effects, we refer the reader to \citet{Freundlich2014}. Another possibility is to consider a more general magnetic field geometry and also to investigate non-axisymmetric perturbation forms. We leave these issues as a matter of study for future works.
\section*{Acknowledgements}
The authors would like to thank Sami Dib for valuable and constructive comments. Also M. Hosseinirad thanks Najme Mohammad-Salehi and Maryam Samadi for useful discussions. S. Abbassi acknowledges support from the International Center for Theoretical Physics (ICTP) for a visit through the regular associateship scheme. We also like to thank the anonymous referee for his/her helpful comments which improved the paper.
\bibliographystyle{mnras}
\bibliography{my_paper} 
\appendix
\section{AD coefficient in dimensionless units}\label{sec:AD_coef}
In our calculations, all quantities are transformed from cgs units to the dimensionless ones. These units are
\be [\rho]=\rho_{c},\ee
\be [t]=\sqrt{4\pi G[\rho]},\ee
\be [\mathbf{u}]=c_{s},\ee
\be [r]=[t][\mathbf{u}],\ee
\be [p]=[\rho][\mathbf{u}]^2,\ee
\be [\psi]=[\mathbf{u}]^2\psi,\ee
\be [\mathbf{B}]=\sqrt{4\pi[\rho]}[\mathbf{u}].\ee
By means of these new units, the unit of $\alpha$ could be expressed as
\begin{equation}
 [\alpha]=\dfrac{1}{4\pi (4\pi G [\rho])^{-1/2}} \frac{4\pi [\rho] [\mathbf{u}]^2}{[\mathbf{u}]^2[\rho]^{3/2}}=\sqrt{4\pi G},
 \label{eq:alpha_dimless}
\end{equation}
which is $\simeq 11.465.$ Moreover, with the help of \cref{eq:eta_A_with_alpha} and \cref{eq:alpha_dimless}, the unit of $\eta_A$ reads
\begin{equation}
 [\eta_A]=\dfrac{1}{4\pi [\alpha][\rho]^{3/2}}.
\end{equation}
This determines $\eta_A$ in dimensionless units as
\begin{equation}
 \eta_A \simeq 0.007 \rho_n^{-3/2}.
\end{equation}
\bsp	
\label{lastpage}
\end{document}